\documentclass[reprint, aps, amsmath, amssymb, twocolumn, superscriptaddress, longbibliography]{revtex4-1}
\usepackage{textcomp}
\usepackage{graphicx}
\usepackage[hidelinks]{hyperref}
\usepackage{xcolor}
\hypersetup{
    colorlinks,
    linkcolor={blue!50!black},
    citecolor={blue!80!black},
    urlcolor= blue
}

\begin{document}

\preprint{APS/123-QED}

\title{Optical nanofiber interferometer and resonator}% Force line breaks with \\

\author{Chengjie Ding}
\affiliation{Laboratoire Kastler Brossel, Sorbonne Universit\'e, CNRS, ENS-PSL Research University, Coll\`ege de France}
\affiliation{State Key Laboratory of Precision Spectroscopy, East China Normal University, Shanghai 200062, China}
\author{Vivien Loo}
\affiliation{Laboratoire Kastler Brossel, Sorbonne Universit\'e, CNRS, ENS-PSL Research University, Coll\`ege de France}
\affiliation{ESPCI Paris, PSL Research University, CNRS, Institut Langevin}
\author{Simon Pigeon}
\author{Romain Gautier}
\author{Maxime Joos}
\affiliation{Laboratoire Kastler Brossel, Sorbonne Universit\'e, CNRS, ENS-PSL Research University, Coll\`ege de France}
\author{E~Wu}
\affiliation{State Key Laboratory of Precision Spectroscopy, East China Normal University, Shanghai 200062, China}
\author{Elisabeth Giacobino}
\author{Alberto Bramati}
\author{Quentin Glorieux}
\email[Corresponding author: ]{quentin.glorieux@lkb.upmc.fr}
\affiliation{Laboratoire Kastler Brossel, Sorbonne Universit\'e, CNRS, ENS-PSL Research University, Coll\`ege de France}

\date{\today}% It is always \today, today,

\begin{abstract}
We report the fabrication and characterization of photonic structures using tapered optical nanofibers.
Thanks to the extension of the evanescent electromagnetic field outside of the nanofiber two types of devices can be built: a ring interferometer and a knot resonator.
We propose a general approach to predict the properties of these structures using the linear coupling theory.
In addition, we describe a new source of birefringence due to the ovalization of a nanofiber under strong bending, known in mechanical engineering as the Brazier effect.
\end{abstract}

\pacs{Valid PACS appear here}% PACS, the Physics and Astronomy
\maketitle

%\tableofcontents

\section{\label{Introduction}Introduction}
Due to their low losses, optical fibers are undoubtedly a medium of choice to transport optical information, making them critical to current telecommunication networks and to the future quantum internet \cite{kimble2008quantum}.
However injecting a specific state of light, for example a single photon in a fiber with a good coupling efficiency is not an easy task. 
%This difficulty has led to the development of embedded light processing units \cite{goure2016optical}. 
A typical way to couple light into a fiber is to place  the emitter directly at one end of a fiber with or without additional optical elements \cite{albrecht2013coupling}.
An alternative approach recently raised significant interest, by injecting light from the side of a fiber \cite{nayak2007optical,le2004atom}. 
Indeed, stretching down the fiber diameter to the wavelength scale allows for a coupling between the fiber guided mode and an emitter in its vicinity \cite{ding2010ultralow}. 
In such a nanofiber, the fundamental propagating mode has a significant evanescent component at the glass/air interface, which allows for interacting with emitters on the surface \cite{yalla2012efficient,schroder2012nanodiamond,vetsch2010optical,nayak2007optical,le2004atom,goban2012demonstration,joos2018polarization}.

Collection efficiency is limited so far to 22.0~$\pm$~4.8\% for a bare nanofiber \cite{yalla2012efficient}. 
Maximizing this coupling is a challenging task as it requires simultaneously a fine-tuning of the fiber size and the largest possible cross-section for the emitter. 
To render this "injection by the side" technique more attractive, the collection efficiency has to be increased.
One approach to do so is to enhance the effective light-matter interaction. 
It is commonly done by reducing the mode volume using confined modes of the electromagnetic field rather than propagating modes. 
It leads, via the Purcell effect,  to an increase of the spontaneous emission within the nanofiber confined mode and therefore to an increase of the emitter-fiber coupling \cite{solano2017optical}.  
A detailed model predicts more than 90\% collection efficiency if one adds an optical cavity of moderate finesse to the nanofiber \cite{solano2017optical}. 
Diverse strategies have been investigated to do so.
 One is to fabricate two mirrors directly in the fiber to add a Fabry--Perot cavity within the nanofiber itself \cite{nayak2014optical}.
This strategy requires advanced nanofabrication methods such as femtosecond laser ablation to modify the fiber index. 
Using a Talbot interferometer, it has been possible to fabricate two fiber Bragg gratings and form an optical cavity with a transmission of 87\% for a finesse of 39 \cite{nayak2014optical}. 
A similar strategy, called nanofiber Bragg cavity, where a focused ion beam mills the nanofiber to create mirrors has shown a Purcell factor and coupling efficiency of 19.1 and 82\% respectively \cite{takashima2016detailed,schell2015highly}. 
Another solution relies on coupling the nanofiber with a whispering gallery mode resonator with very high quality factor up to $10^9$ \cite{cai2000observation,aoki2006observation}. 
With this strategy, at the difference of previous ones, the cavity is exterior to the nanofiber.

\begin{figure}[htbp]
\centering\includegraphics[width=1\linewidth]{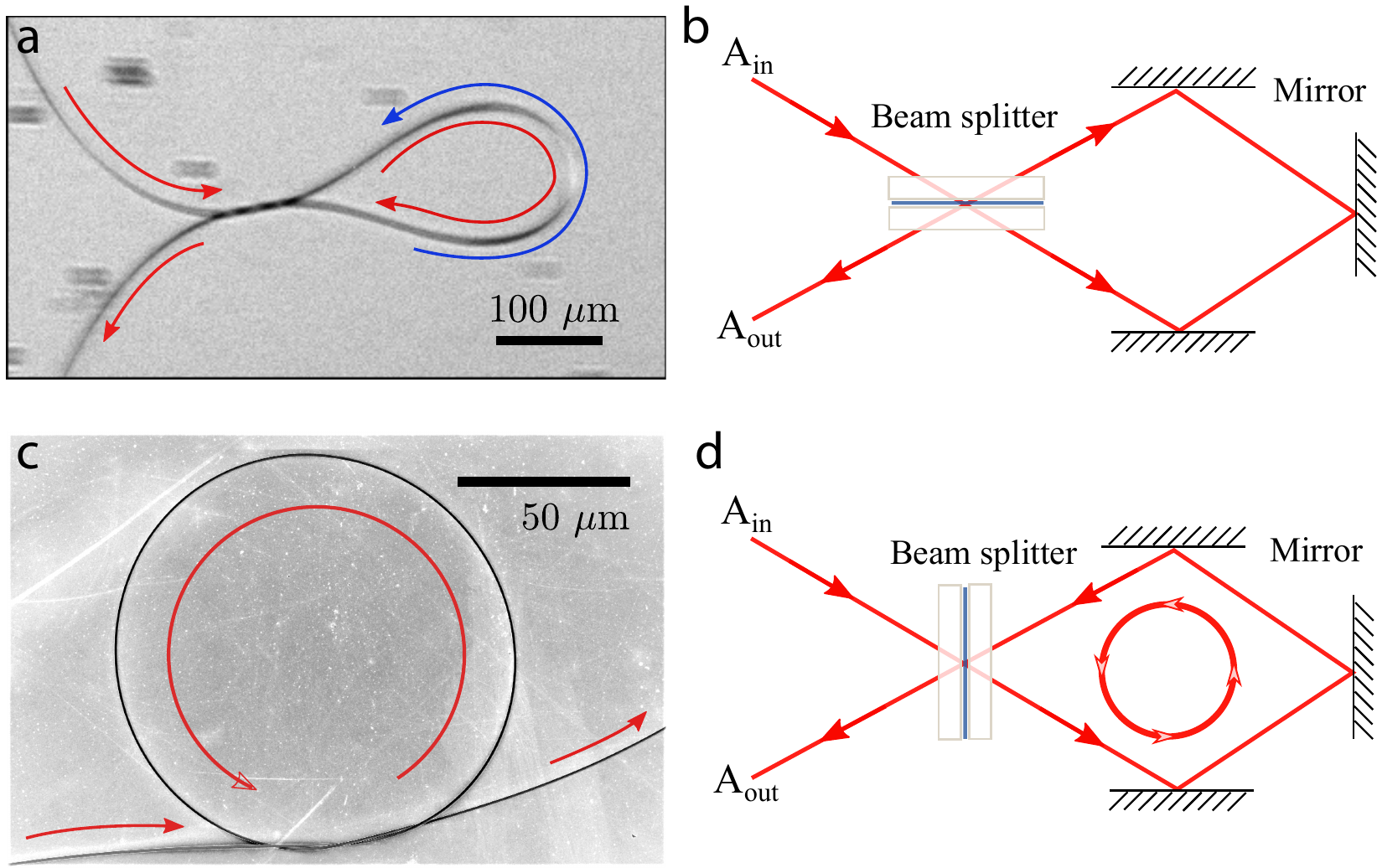}
\caption{Optical nanofiber structures.
a- Nanofiber twisted loop: optical microscope image. 
b, Sagnac interferometer equivalent optical setup: the light emerging from the port $A_{\text{out}}$ consists of two reflections or two transmissions of the light from incident port $A_{\text{in}}$ through the beamsplitter. 
c- Nanofiber knotted loop: scanning electron microscope image. 
d- Fabry-Perot ring resonator equivalent optical setup: light coming from $A_{\text{in}}$ that is not directly reflected to $A_{\text{out}}$ by the beam splitter, is trapped in the cavity formed by the beam splitter and the mirrors.}\label{figure1}
\end{figure}

In this article, we study an alternative approach, particularly interesting because it does not involve nanofabrication capabilities. 
The idea is to loop the nanofiber in order to directly create a cavity thanks to the evanescent coupling. 
Such cavities have been obtained in the telecom range at 1.5 microns using microfiber with a finesse up to 20 \cite{sumetsky2005optical,jiang2006demonstration}. 
Here we demonstrate the first experimental implementation of this approach in the visible range using a fiber tapered to the nanoscale while maintaining a similar finesse.
Moreover, using the linear coupling theory, we present a generalized model for two complementary geometries: twisted and knotted loops, illustrating the crucial role of the topology of the loop formed.
While the twisted loop is found to work as a  Sagnac interferometer, the nanofiber knot  behaves as a Fabry--Perot micro-resonator. 

In addition, we report here a novel source of birefringence for nanoscale tapered fiber. 
In such micron-size structures the nanofiber region is put under strong bending constraints and therefore this induces an ovalization of its transverse section, known as the Brazier effect in mechanical engineering \cite{tatting1997brazier}.
We have estimated that this ovalization leads to a substantial difference in the effective refractive index similar to the the well-known stress induced birefringence \cite{okamoto2006fundamentals}.

\section{\label{sec:level1}Effective coupling theory approach}

The manufacturing of nanofibers is a well-controlled process, and it is possible to fabricate fibers with a diameter down to 200 nm \cite{ward2014contributed,ding2010ultralow}. 
At this size, only the core of the fiber remains, and the surrounding air acts as a cladding. Consequently, there is a strong evanescent field extending around the surface of the nanofiber.
The fundamental mode does not correspond anymore to the standard linearly polarized mode LP$_{01}$. Nevertheless, using Maxwell's equations the correct propagating mode profile can be precisely characterized \cite{le2004field}.
We will consider single mode air-cladding nanofibers only, that is, nanofibers in which the fundamental mode HE$_{11}$ is the only propagating solution\cite{le2004field}. This is the case if the normalized frequency $V$ with $V\equiv ka\sqrt{n^2-1}$ is lower than the cutoff normalized frequency $V_{c}=2.405$, where $k$ is the wavevector, $a$ is the fiber radius, and $n$ is the fiber index.

We have bent and twisted manually such nanofibers with great care to realize two miniaturized optical setups: a fiber loop and a fiber knot (Fig. \ref{figure1}).
The common feature of these two structures is that they both present a section where the two parts of the nanofiber touch each other as shown on Fig. \ref{figure1}. 
However, fiber knots and fiber loops are topologically distinct, as it will be detailed later.

\begin{figure}[htbp]
\centering\includegraphics[width=0.5\linewidth]{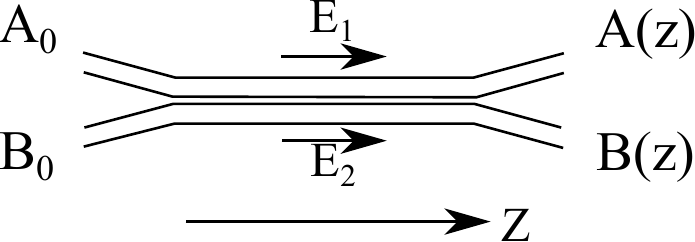}
\caption{Schematic of codirectional couplers. $A_0$ and $B_0$ are the amplitude of electric fields at the input of the two fibers. The outputs of each fiber are labelled $A(z)$ and $B(z)$.}
\label{figure0}
\end{figure}

Given the strong evanescent field of the propagating mode, the contact between different nanofiber regions leads to a coupling between the HE$_{11}$ propagating modes. 
To model this coupling, let us consider two parallel nanofibers nearby to each other, as represented in Fig.~\ref{figure0}.
 In this scheme, the two fibers exchange energy in the contact region of length $z$ with coupling coefficient $\kappa$.  
Therefore, given the input amplitudes $A_0$ and $B_0$
the output amplitudes are:
\begin{align}
 A(z) &= A_0\cos(\kappa z)  - i B_0\sin(\kappa z)  \label{Az}\\
 B(z) &= B_0 \cos(\kappa z)  - i A_0 \sin(\kappa z)\; . \label{Bz}
\end{align}

The coupling coefficient $\kappa$ depends on the overlap of the coupled modes \cite{okamoto2006fundamentals}:
\begin{equation}
 \kappa=\frac{\omega \epsilon_0 \iint (n_{\text{air}}^2-n_{\text{silica}}^2) \mathbf{E}_1^* \cdot \mathbf{E}_2 \ \text{d}x \text{d}y}{\iint \mathbf{u}_z \cdot (\mathbf{E}_1^* \times \mathbf{H}_1 + \mathbf{E}_1 \times \mathbf{H}_1^*) \  \text{d}x \text{d}y} \label{k}
\end{equation}
where $\mathbf{E}_i$ and $\mathbf{H}_i$ are respectively the electric and magnetic components of the modes propagating in the nanofiber labeled $i$, $\mathbf{u}_z$ is the unitary vector directed toward the propagation direction and $\omega$ the field frequency.

According to Eqs. (\ref{Az}) and (\ref{Bz}), we can regard the light going through the fiber 1 as transmitted, and the light going from fiber 1 to fiber 2 as reflected. 
The system acts as a beam-splitter of transmission coefficient $t=\cos(\kappa z)$, and reflection coefficient $r=-i\sin(\kappa z)$. 
For example, when the system has an input $A_0=1$ and $B_0=0$, the output intensities appear to be $|A(z)|^2=\cos^2(\kappa z)$ and $|B(z)|^2=\sin^2(\kappa z)$, and complete power transfer occurs when $\kappa z=(2p+1) \pi / 2$, $p$ being an integer.
Consequently the quantity $ \pi / 2\kappa$ is equivalent to a coupling length.

Interestingly the orientation of the effective beam splitter depends on the topology of  the structure. 
In the case of the twisted loop represented in Fig.~\ref{figure1}-b, the beam splitter is equivalent to a Sagnac interferometer allowing for only one lap in the structure. 
Whereas in the case of the knot (Fig.~\ref{figure1}-d), the effective beam splitter allows for multiple laps inside the setup and therefore is equivalent to a ring resonator.

\begin{figure}[t]
\centering\includegraphics[width=\linewidth]{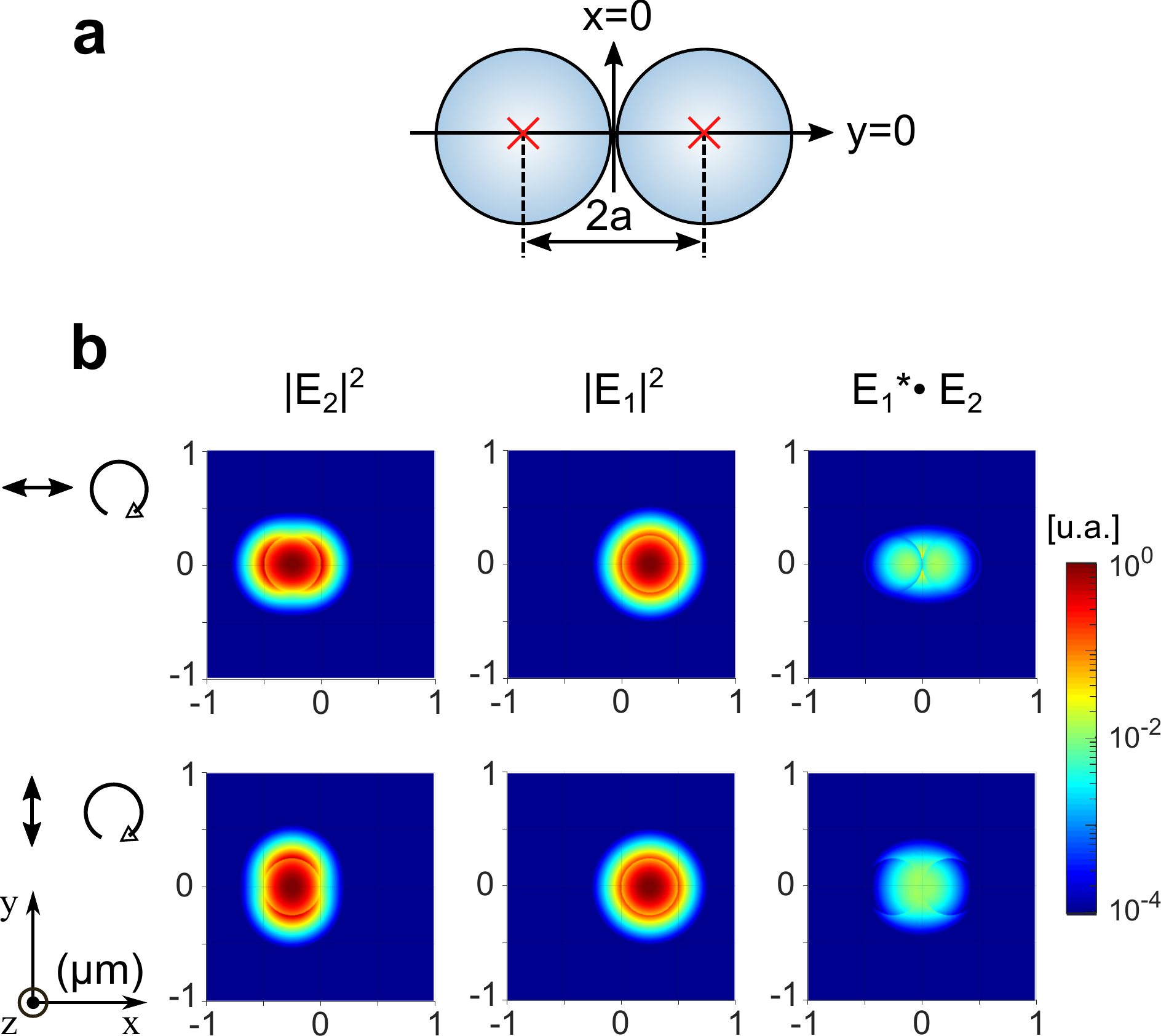}
\caption{Field distribution and overlap for two adjacent nanofibers (fiber radius $a$ = 250 nm). a, The centers of two nanofibers are located on axis y=0. 
Fiber 1 is located at position ($a$,0), and fiber 2 is located at ($-a$,0). 
b, The light in the fiber 2 corresponds to an HE$_{11}$ and  linearly polarized as visible in the left panel (upper left panel for the horizontally polarized and lower left panel for vertically polarized). 
The field is circularly polarized in fiber 1 as visible in the central panels. 
Right panels represent the overlap between the two fields directly related to the coupling strength $\kappa$ given in Eq.~(\ref{k}).
}\label{figure2}
\end{figure}

As mentioned above, we place ourselves in conditions under which only one propagating mode exists: HE$_{11}$.
To estimate the coupling coefficient $\kappa$, we computed Eq. (\ref{k}) using the exact profile of modes HE$_{11}$ \cite{le2004field}. 
In order to study the coupling coefficient dependence on the polarization, we assume that nanofibers are identical and in contact at (0,0) on $x-y$ plane, as shown in Fig.~\ref{figure2}-a, and that the light is linearly polarized in one fiber, whereas it is circularly polarized in the other one.

\begin{figure}[htbp]
\centering\includegraphics[width=0.95\linewidth]{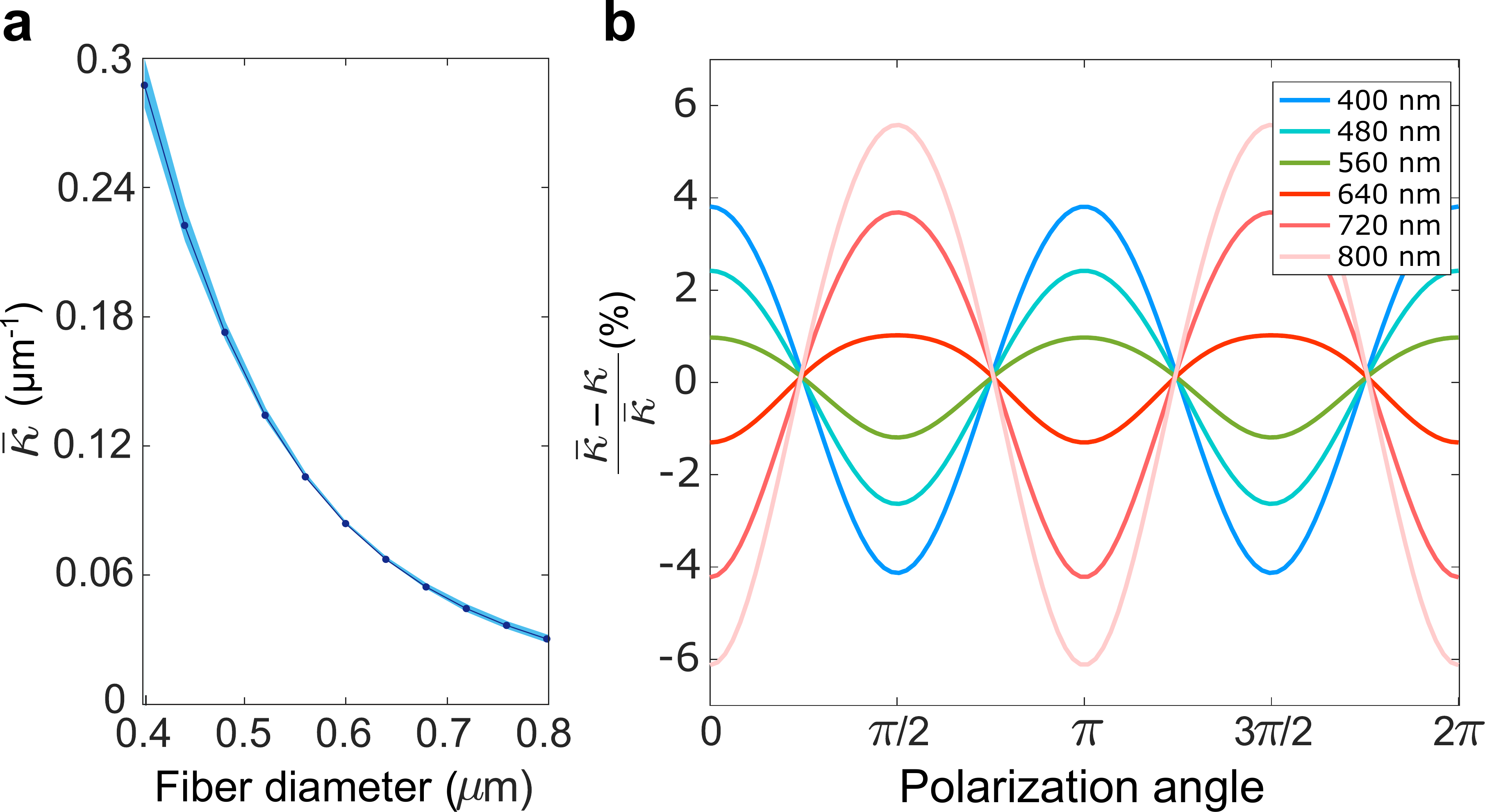}
\caption{a- Coupling coefficient $\kappa$ (at 800~nm) between nanofibers with varying diameter and averaged over the polarization degree of freedom. 
The colored area refers to the amplitude of the variation with the polarization. 
b- Relative change of the coupling coefficient as a function of the polarization angle with a fiber diameter varying from 500 nm to 900 nm.}\label{figure3}
\end{figure}

The field density in the transverse section in this configuration is presented in Fig.~\ref{figure2}-b.
We numerically calculated the coupling strength as a function of the fiber diameter averaged over the polarization degree of freedom $\bar\kappa = \langle \kappa \rangle_{\varphi}$, where $\varphi$ is the angle of the polarization vector with the $x$ axis.
Results are presented in Fig.~\ref{figure3}-a.
We see that $\bar\kappa$ decays exponentially with the fiber diameter. 
The larger the fibers are, the smaller their evanescent part of the field is. 
This decay is exponential so it is for the overlap of the fields. This strong dependency illustrates well the general interest to work with fiber at subwavelength scale rather than micrometric scale. 
Focusing now on the polarization dependency of the coupling strength we show two cases in Fig.~\ref{figure2}-b : 
(i) the polarization of the linearly polarized field in fiber 2 is along the $x$ axis (along the direction that connects the centers of the two fibers)  ($\varphi=0$) (upper panels of Fig.~\ref{figure2}-b) and 
(ii) the polarization of the linearly polarized field in fiber 2 is normal to the $x$ axis  ($\varphi=\pi/2$) (lower panels of Fig.~\ref{figure2}-b). 
For both cases the field density is presented for the linearly polarized field in fiber 2 (left panels), for the circularly polarized field in fiber 1 (central panels) and for their overlap $\mathbf{E}_1^*\cdot\mathbf{E}_2$ appearing in Eq. (\ref{k}), where  $\mathbf{E}_1^* \cdot \mathbf{E}_2=\mathbf{E}_{1x}^*\cdot \mathbf{E}_{2x}+\mathbf{E}_{1y}^*\cdot \mathbf{E}_{2y}+\mathbf{E}_{1z}^*\cdot \mathbf{E}_{2z}$.
The results shown are for a fiber diameter of 500 nm and wavelength of 800 nm.
 We see that even if the overlap intensity distribution is different from one case to the other,  their average magnitude and then the coupling coefficient are similar in the two cases as shown in Fig.\ref{figure2}-b.  
 
Actually, the coupling coefficient $\kappa$ is found to be only slightly dependent on the polarization. 
This variation depends on the fiber diameter, as visible in the b panel of Fig.~\ref{figure3} but leads to a marginal relative change.
 In the realization of the nanofiber twisted loop below, we use a fiber diameter of 500 nm for which the relative change is estimated to be less than $\pm 3\%$. 
This is illustrated in Fig.~\ref{figure3}-a by the colored region surrounding the mean coupling strength $\bar\kappa$, corresponding to the amplitude of the variation with respect to the polarization. 
Given that, we can reasonably neglect the effect due to polarization and approximate $\kappa\approx\bar\kappa$.

\section{\label{sec:level2}Nanofiber interferometer}

We realized an optical nanofiber by pulling a commercial single-mode fiber to reach a 500~nm diameter over a length of 1~mm following \cite{hoffman2014ultrahigh}.
The transition between the commercial single mode fiber and the single mode nanofiber is adiabatic and its transmission is over 95\% \cite{joos2019arxiv}.

To make a twisted loop structure presented in Fig.\ref{figure1}-a,  we first make a ring in the nanofiber region. 
Then, by fixing one side of the nanofiber, and rotating the other side, we can slowly increase the length of the entwined part. 
This is a well-known mechanical phenomenon studied in many contexts \cite{goriely1998nonlinear}.
Increased torsion will reduce the size of the loop and bend it locally.
At some point the bending exceeds the fiber tolerance and it breaks. 
In the experiment we carefully choose to remain below this threshold.

With this geometry, the system corresponds to a Sagnac interferometer. 
As represented in Fig.\ref{figure1}-b, light propagating towards the loop finds two counter-propagating optical paths.
After the entwined region, part of the light is transferred into the clockwise path (red arrow) with a coefficient of $r$, whereas the remaining propagates along the anti-clockwise path with a coefficient of $t$. 
This two beams propagate separately accumulating a phase of $e^{i \beta L_r}$, where $\beta$ is the propagation constant, and $L_r$ is the length of the ring.
 Then, both paths interfere back into the entwined region, which acts as a beam-splitter, as represented in Fig.\ref{figure1}.b.  
The amplitude of output electromagnetic field can be written as:
\begin{equation}
 A_{\text{out}}=[r^2 +t^2]e^{i\beta L_r}A_0,
\end{equation}
with $t=\cos(\kappa z)$, and $r=-i\sin(\kappa z)$, as shown above, 
which leads to the following transmittance for the device:
\begin{equation}
 T_{\text{int}}=\left|\frac {A_{\text{out}}}{A_0}\right|^2=\left|r^2 +t^2\right|^2.
\end{equation}
As mentioned above, we control the length of the coupling region by varying the torsion applied on the nanofiber, which ultimately tunes the reflection and transmission coefficients of our device.
Applied mechanical stress will be translated into an optical response, from reflective to transmissive.

\begin{figure}[htbp]
\centering\includegraphics[width=0.95\linewidth]{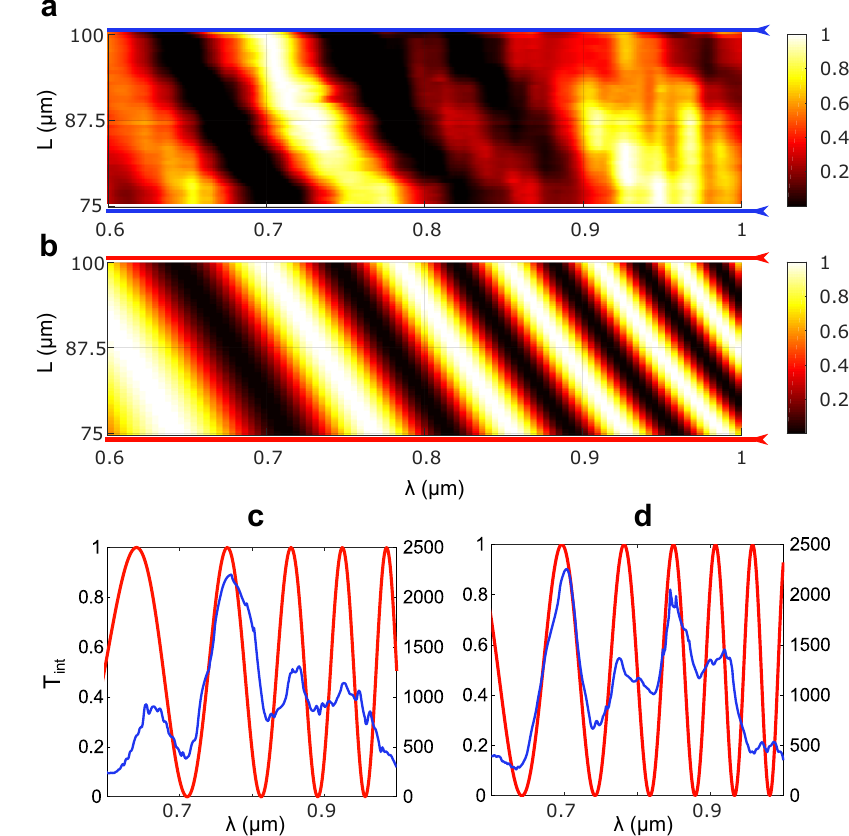}
\caption{Transmission spectrum through twisted loop structure. 
a and b- Experimental data and numerical simulation of the transmission spectrum of broad light source through the nanofiber interferometer as a function of the wavelength and coupling distance.
c and d- Experimental data (blue) and numerical simulation (red)  of the transmission spectrum for an entwined length of 75 $\mu$m (panel c) and 100 $\mu$m (panel d) respectively. 
The experiment data shown are smoothed using rloess methods with a span of 1\%.}\label{figure4}
\end{figure}

To probe the system transmittance we use a fiber-coupled linearly polarized super-continuous white laser (NKT Photonics SuperK COMPACT). 
The laser beam is coupled into the single mode fiber (SM800) with the twisted nanofiber region in the middle. %The polarization entering the nanofiber structure can be adjusted with a 3-Paddle Polarization Controllers. 
The output signal is sent to a spectrometer.
Fig.~\ref{figure4}-a shows a map of the transmission spectrum through the nanofiber twisted loop structure as a function of the entwined region size.
%Latter is directly estimate thanks to a confocal microscopy image as the one presented in Fig.~\ref{figure5}.a. 
To understand the spectrum obtained for a given entwined region length, also presented in Figs.~\ref{figure4}-c and d in blue, one has to note that when the fiber diameter is fixed, the extension of the evanescent part of the field increases with the wavelength. Accordingly the coupling strength $\kappa$ and so the effective reflection coefficient $r$  change too as shown by the red curves in Fig.~\ref{figure4}-c and d. In consequence, the spectrum for a fixed coupling length leads to the interference pattern visible in Fig.~\ref{figure4} and agrees with our description of the device as an interferometer.  
Moreover, increasing the coupling length leads to a shift of the interference to smaller wavelengths.
In Fig.~\ref{figure4}-b we represent the same map as in Fig.~\ref{figure4}-a, calculated from Eq. \ref{k}.
Despite the variability of many experimental parameters, our theoretical model shows good agreement with the experimental data,
This agreement can be verified quantitatively in panel c and d, where we show the measured (blue) and calculated (red) spectrum for a coupling length of respectively 75 $\mu$m and 100 $\mu$m with no adjustable parameters.

\section{\label{sec:Nanofiberresonator}Nanofiber resonator}

Optical ring resonators can have many applications, such as optical add-drop filters, modulation, switching and dispersion compensation devices. 
To fabricate the knot structure as presented in Fig.~\ref{figure1}-c, we carefully made a large knot centered on the nanofiber region and by precise control of the spacing between the two displacement platforms used to pull the fiber, we can decrease the diameter of the fiber to tens of micrometers ($\approx 30~\mu$m). 
As in the case of the twisted loop, the knotted loop induced an important bending of the nanofiber.
Accurate control of the size of the knot allows us to avoid breaking it. 

With this geometry, the system acts as a resonator.
As represented in Fig.~\ref{figure1}-d, the light injected in the device will either be directly reflected to the output with a coefficient $r$, or transmitted in the loop with a coefficient $t$.
In contrast with the twisted loop, there is only one optical path within the loop.
Moreover, the light circulating inside the knot will split again everytime it passes through the coupling region: some will go to the output, the rest will stay inside the knot. 
We represent in Fig.~\ref{figure1}-d the corresponding optical setup. 
It is remarkable that the change of the topology of the loop formed, twisted or knotted, completely changes the behavior of the device. 
Schematically, passing from one device to the other one is equivalent to rotating by 90$^\circ$ the beam splitter mimicking the fiber coupling region as presented in Fig.~\ref{figure1}-b and Fig.~\ref{figure1}-d. 

\begin{figure}[htbp]
%\graphicspath{ {E:/Pictures/} }
\centering\includegraphics[width=0.95\linewidth]{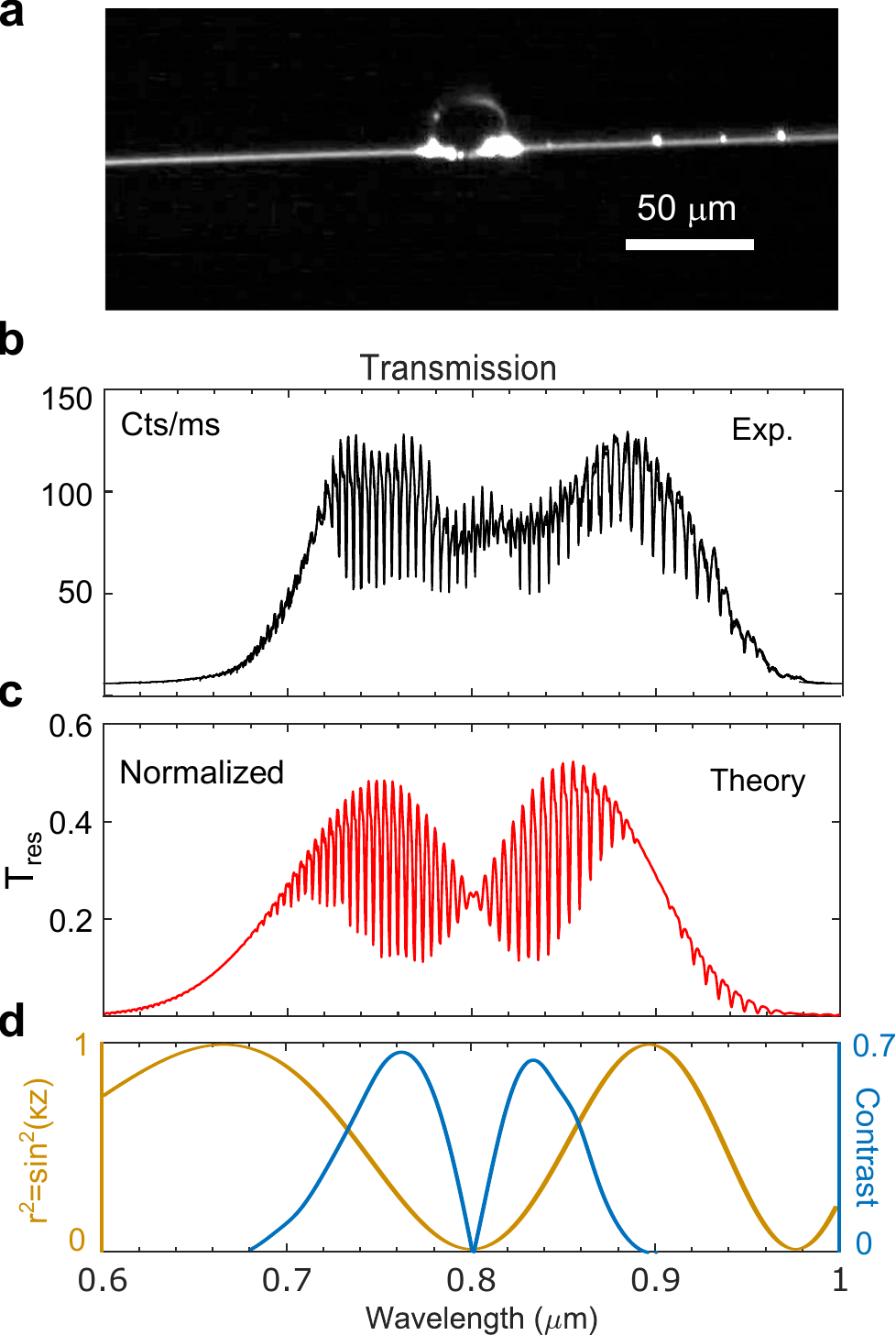}
\caption{Transmission spectrum through a knot structure.
a- Image of the nanofiber knot recorded by an optical microscope.
b- Experimental transmission spectrum of a broad light source through the nanofiber resonator as a function of the wavelength. 
c- Calculated transmittance of the devices as given in eq. (\ref{tres}). 
d- Left axis is the reflectance of the contact zone and right axis is the visibility of resonance between wavelength 0.68 $\mu m$ and 0.9 $\mu m$. 
We measure $\Delta\upsilon_{\text{FSR}}$=2.214 THz.}
\label{figure5}
\end{figure}

Along the propagation in a loop, the field undergoes losses with a rate $\rho$ due to scattering. 
In Fig. \ref{figure5}-a we present an optical microscopy image of the knotted loop when light is propagating on the fiber. 
Bright spots on the fiber are due to the scattering of imperfections on the fiber surface.
Given the significant evanescent component of the field, any defects located close to the surface will strongly scatter the propagating light.
However, most of the losses come from the knotted region itself as visible in Fig. \ref{figure5}-a. 
They would drastically be reduced in a clean room environment.
Indeed, when we fabricate the knot manually, the knot was gradually tightened into small size. Thus, the entwined region sweeps several centimeters of fiber.
Impurities on the surface are blocked by the knot and inevitably accumulate there. 

After one lap in the loop we have $B'_0=(1-\rho)e^{i\beta L_k}t^2B_0$, where $L_k$ is the length of the ring. 
Assuming the reflection coefficient $r=- i \sin(\kappa z)$ and transmission coefficient $t=\cos(\kappa z)$, then we get the equation giving the amplitude of the electromagnetic field at the output:
\begin{equation}
\begin{aligned}
A_{\text{out}}=A_0[ r + (1-\rho)t^2 e^{i \beta L_k}  + (1-\rho)^2t^2 r e^{2i \beta L_k} \\+ (1-\rho)^3t^2 r^2 e^{3i \beta L_k} + \cdots ]
\end{aligned}
\end{equation}
leading to the following transmittance $ T_{\text{res}}$ of the device:
\begin{equation}
 T_{\text{res}}=\left|\frac {A_{\text{out}}}{ A_0}\right|^2=\left|r + \frac {(1-\rho)t^2e^{i \beta L_k} }{1- (1-\rho)re^{i \beta L_k}}\right|^2 \label{tres}
\end{equation}

In contrast to the nanofiber interferometer case, the size of the loop plays a major role here as it dictates the amount of phase accumulated after one lap in the device.
Optical resonances appear when the light traversing the loop accumulates a phase integer multiple of $2\pi$.  

To characterize the system we use the same setup as for the twisted loop.
Fig. \ref{figure5}-b is the transmission spectrum of the knot for a given diameter. Within the bandwith of the SM800 fiber ($\approx$ 700 nm to $\approx$ 950 nm), many fine peaks are observed, revealing the resonant wavelengths.

In Fig. \ref{figure5}-c, we present the calculated transmission spectrum, which agrees well with the experimental data, for its three main features.
Firstly, the wide spectral Gaussian envelop; this is measured beforehand by recording the spectrum of our laser transmitted through a fiber without the loop.
Secondly, the fine peaks; they exhibits matched free spectral range (FSR) and amplitudes.
Finally, the larger scale contrast modulation ; its maximum around 750 nm and 850 nm are faithfully reproduced.
To explain this contrast modulation, we represent in Fig. \ref{figure5}-d the reflectance of the entwined region (i.e. without knot) calculated for the same condition. 
As observed previously, it depends strongly on the wavelength. 
For instance, at $\lambda\approx 800$ nm, the reflectance $|r|^2$ is zero; it corresponds to a scenario without beamsplitter.
As 100\% of the light leaves the ring after one lap, there cannot be interferences, and the resonance peaks contrast vanishes accordingly. 
Similarly when $r^2\sim 1$, around 900 nm and 670 nm, the entwined region acts as a simple mirror instead of a beam-splitter, and the resonance peaks fade as well, since no light gets inside the ring.

The length of the ring $L_k$ determines the interval between the peaks in the spectrum known as FSR and given by $\Delta\upsilon_{\text{FSR}}=c/(n_{\text{eff}}L_k)$, where $n_{\text{eff}}$ is the effective refractive index of air clad optical fiber. 
The analysis of the spectrum gives $\Delta\upsilon_{\text{FSR}}$=2.21 THz, which corresponds to a cavity length of about 108~$\mu$m. 
The finesse varies slightly along the spectrum reaching 8 from 820 nm to 860 nm. 
In this range, we measure a quality factor of 1300. 
It agrees  well with the calculation which predicted a $\Delta\upsilon_{\text{FSR}}$ of 2.136 THz, a finesse of 7.5 and a quality factor of 1100. 
These calculations have been done with an estimate of 35\% losses, extracted from experimental data.  
In the next section, we push forward the analysis of the spectrum and we identify an original birefringence effect.

\subsection*{Birefringence induced by ovalization under bending}
\begin{figure}[htbp]
\centering\includegraphics[width=\linewidth]{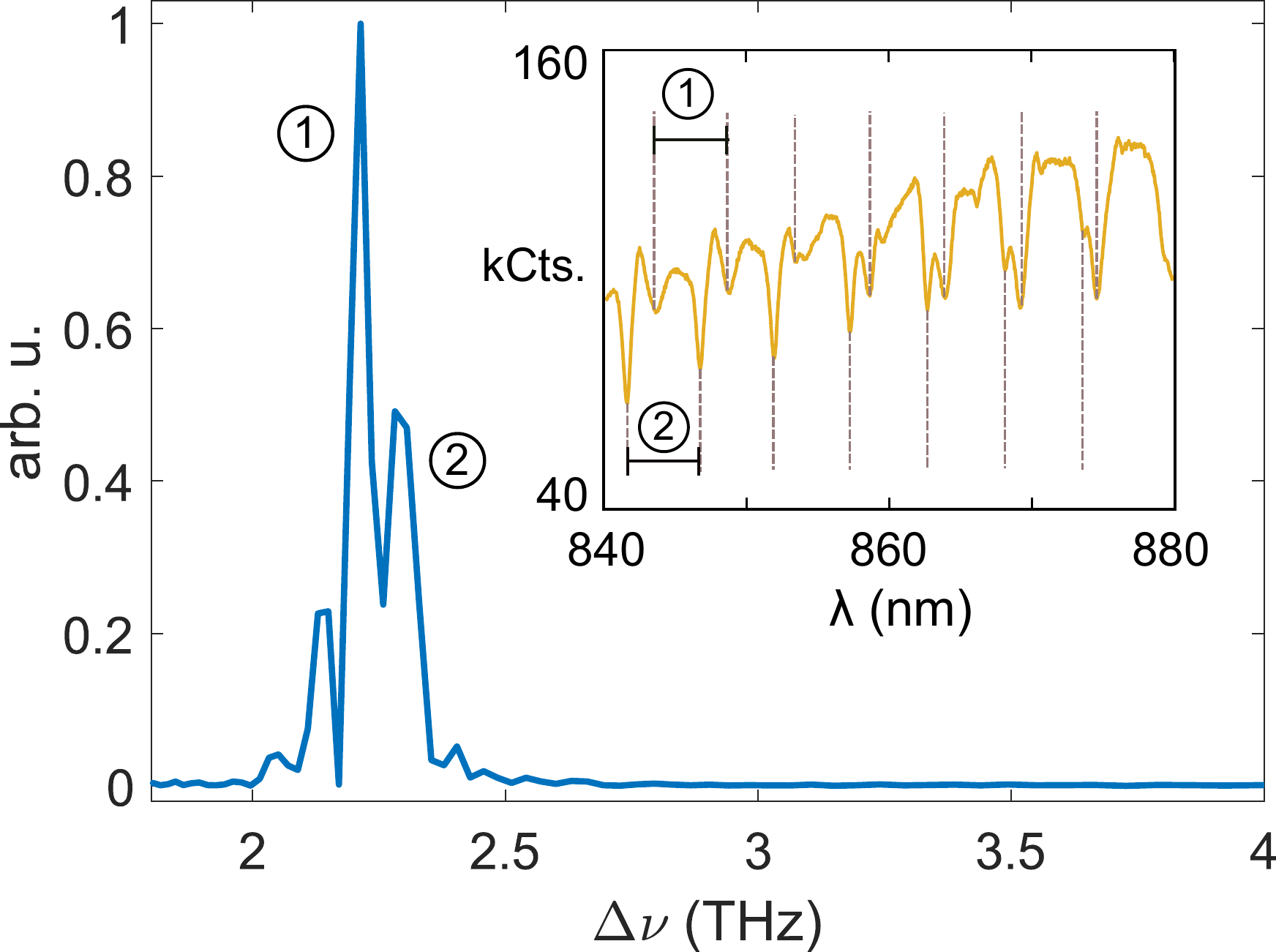}
\caption{Blue: Fourier transform of experimental transmission spectrum. 
Inset yellow: zoom of the Fig 6-b.
Two different sets of peaks can be seen.
The corresponding $\Delta\upsilon_{\text{FSR}}$ are 2.21 THz and 2.28 THz (peaks location in the blue curve).
This splitting corresponds to different polarization modes undergoing birefringence along the propagation in the structure.}\label{figure6}
\end{figure}

A keener look at the data reveals that two different resonance modes (see insert curve in Fig.~\ref{figure6}) contribute to the spectrum, otherwise we would observe regularly spaced peaks.
The FSR for the different modes is slightly offset.
When the peaks positions of the two modes are staggered, we can clearly see the two discrete peaks.
To identify the two values of FSR, we took the Fourier transform of the spectrum, which is plotted in Fig.\ref{figure6}. 
There, we clearly see two distinct resonant contributions labeled 1 and 2. 
This shift between the FSR values corresponds to a relative variation of the indices between the two propagating modes of $\Delta=(n_1-n_2)/n_1=3.1\%$, with $n_1$ and $n_2$ the effective refractive indices of both modes.

To determine the origin of this lifting of degeneracy, we will now focus on the mechanical properties of the knot.
As for the twisted loop we have here a nanofiber under strong bending constraints.
This bending implies stress and related mechanical effect that we assume at the origin of the mode splitting. 
We investigate two possible sources of birefringence: (i) stress-induced birefringence ($B_s$) and (ii) ovalization-induced birefringence ($B_o$) \cite{okamoto2006fundamentals}. 
Stress-induced birefringence is a well-known phenomenon in optical fiber \cite{ulrich1980bending} whereas ovalization-induced  birefringence is directly linked to the diameter of the fiber considered here.

Ovalization of an elastic rod is a well-known phenomenon in mechanical engineering \cite{wierzbicki1997simplified}, and can be commonly experienced when bending an elastic tube. 
The perfect circular section of the tube will change due to the bending to an oval section with its short axis along the direction of the bending. 
Such a mechanism is commonly neglected in optical fibers given the rigidity of standard fibers in which a bending force will lead to stress-induced birefringence and break the fiber before leading to significant ovalization of the section. 
However, in our case, with fiber of nanometric diameter, the mechanical properties are very different.
Moreover, the electromagnetic field is significantly less sensitive to stress, since it is largely localized outside the fiber. 
The stress only affects the part of the mode confined inside the silica. 
We evaluate the relative change of indices due to stress to be of $\Delta\approx1.7\%$. This is not enough to explain the different values of FSR we observed, and leads us to consider ovalization, or Brazier effect, as a significant factor in the degeneracy lifting.   

\begin{figure}[htbp]
\centering\includegraphics[width=\linewidth]{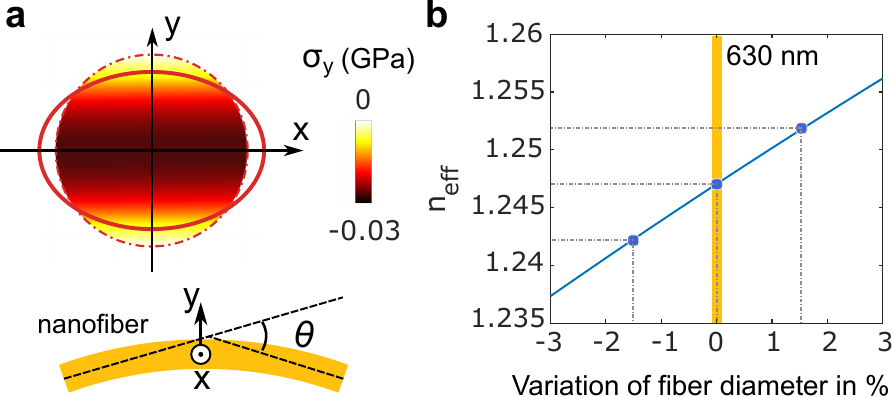}
\caption{Brazier effect on a bent optical fiber. 
a- Schematic of the fiber cross-section ($x-y$ plane) ovalization. 
$\sigma_y$ is the $y$ component of the bending induced stress, represented with the color scale within the fiber cross-section.
$\theta$ is the angle of rotation of the fiber ends.
b- Relative refractive index variation as a function of the fiber diameter change in percent compared with an initial diameter of 630 nm.}\label{figure8}
\end{figure}

We observe that ovalization or Brazier effect also participates in the birefringence. 
Brazier effect is a mechanical deformation that causes the ovalization of the cross-section of a bent tube.
 As shown in the Fig.~\ref{figure8}-a, $\sigma_y$ is defined as the $y$ component of the bending induced stress. The color in the fiber cross-section shows the value of $\sigma_y$ given by \cite{ulrich1980bending}:
 \begin{equation}
    \sigma_y = K^2 (E/2) (x^2-a^2)
\end{equation}
where $E$ is the Young's modulus of silica, $K=1/R$ is the curvature of the longitudinal axis, and $a$ is the fiber radius. 
 The value of $\sigma_y$ varies along the $y$ axis from 0 GPa at the fiber surface to -0.03 GPa at $y=0$ (corresponding to the compression stress), which gives an increased circumferential strain at fiber surface along the $x$ axis and a reduced circumferential strain at fiber surface along the $y$ axis \cite{wang2018cross}. 
 This effect, named as Brazier effect, caused the ovalization in the plane perpendicular to the bending axis. 
The relative displacement of two axes can be approximated as \cite{wierzbicki1997simplified}:  
\begin{equation}
    \overline{\delta}=0.553 aK
\end{equation}
where $a$ is the initial fiber radius before bending. 
Taking into account the measured loop radius ($R\approx11 \mu m$) and the fiber diameter  we found a reduction of the fiber diameter of  1.5\% parallel to the bending direction and an increase of 1.5\% in the normal direction. 
In contrast to standard fibers, this small change in the geometry will have strong impact given the transverse distribution of the field. 
A difference of 3\% of the radius, leads to a difference of effective refractive index of $(n_{+1.5\%}-n_{-1.5\%})/n=0.8\%$, as shown in Fig.\ref{figure8}-b.
 The joint effect of stress and ovalization give a birefringence of 2.5\%, which is in reasonable agreement with the observed value ($\Delta=3.1\%$). 
The small discrepancy between these two values is likely due to the fact that  the loop is not perfectly circular (as visible in Fig.\ref{figure5}-a ) and therefore leads to a non-homogeneous ovalization and stress effect along the ring.
\section{\label{sec:conclusion}Conclusion}
In this paper, we have reported the fabrication and characterization of two optical devices based on looped nanometrical optical fibers. 
The nature of the devices changes with the topology of the loop. 
The entwined part of the loop can be treated with the coupled mode theory and can be seen as a tunable beam splitter. 
We showed that a twisted loop creates a Sagnac interferometer, of which the dephasing is tuned by the torsion applied on the nanofiber. 
The second device is a cavity, simply made out of a nanofiber knot. 
We analyzed its spectral response and found a finesse of 8 and a quality factor of 1100. 
Unlike in common resonators, the coupling efficency into the cavity strongly depends on the wavelength, which modulates the visibility of its resonances. 
It is reproduced accurately by our theoretical model.
Both setups, Sagnac interferometer and Fabry--Perot resonator are essential to photonics applications and optics in general. 
Their miniaturized versions presented here pave the way toward their integration in photonic circuits. 
 A refined analysis of the cavity spectrum revealed that birefringence of a bent nanofiber is also affected by ovalization of its profile, and not only by stress as a normal fiber would be.
With both devices, we showed how sensitive nanofibers are to mechanical constraints. In this regard, they could open an exciting playground in optomechanics.
\section{\label{sec:acknowledge}Acknowledgments}
This work has been supported by the C-FLigHT ANR project, Emergences Ville de Paris Nano2, Caiyuanpei Programm.
C.D. is supported by the CSC scholarship.

\bibliography{main.bbl}
\bibliographystyle{unsrt}
\end{document}